\documentclass[11pt]{article}
\usepackage{amsmath}
\usepackage{amssymb}
\usepackage{latexsym}
\usepackage{epsfig}

\newcommand{\be}{\begin{equation}}
\newcommand{\ee}{\end{equation}}
\newcommand{\refb}[1]{(\ref{#1})}

\newcommand{\sectiono}[1]{\section{#1}\setcounter{equation}{0}}

\begin{document}

{}~
\hfill\vbox{\hbox{hep-th/0406212}\hbox{MIT-CTP-3506}
}\break

\vskip 3.8cm

\centerline{\Large \bf Heterotic String Field Theory}

\vspace*{10.0ex}

\centerline{\large \rm Yuji Okawa and Barton Zwiebach}

\vspace*{8.0ex}

\centerline{\large \it Center for Theoretical Physics}

\centerline{\large \it
Massachusetts Institute of Technology}

\centerline{\large \it Cambridge,
MA 02139, USA}
\vspace*{1.0ex}

\centerline{okawa@lns.mit.edu, zwiebach@lns.mit.edu}

\vspace*{10.0ex}

\centerline{\bf Abstract}
\bigskip
\smallskip

We construct the Neveu-Schwarz sector
of heterotic string field theory
using the large Hilbert space of the superghosts
and the multi-string products of bosonic closed string field theory.
No picture-changing operators are required
as in Wess-Zumino-Witten-like open superstring field theory.
The action exhibits a novel kind of nonpolynomiality:
in addition to terms necessary
to cover missing regions of moduli spaces,
new terms arise from
the boundary of the missing regions and its subspaces.
We determine the action up to quintic order
and a subset of terms to all orders.

\vfill \eject

\baselineskip=16pt


\sectiono{Introduction} \label{s1}

String field theory is one possible approach
to the construction of a nonperturbative formulation
of string theory.
There are two known Lorentz-covariant
bosonic string field theories.
One is an open string field theory~\cite{Witten:1985cc}
with a Chern-Simons-like cubic action,
and the other one is
a closed string field theory~\cite{Zwiebach:1992ie,Saadi:tb}
with nonpolynomial interactions which are necessary
to cover the moduli spaces of punctured spheres.
There is also an open-closed bosonic string field
theory~\cite{Zwiebach:1997fe}, and there is
a conjectural vacuum string field theory~\cite{vsft}.

The construction of superstring field theory
is complicated because of superghost pictures.
For the Neveu-Schwarz (NS) sector of open superstrings,
the difficulties have been  resolved
by working in the large Hilbert space that includes
the zero mode of the fermionic superghost $\xi$.
The resulting Wess-Zumino-Witten-like (WZW-like)
action~\cite{Berkovits:1995ab}
contains an infinite number of regular contact terms.
The great advantage of this formulation is that
no picture-changing operators appear in the action.
It is also possible to write heterotic
and, perhaps, type II string field theories
if we allow the insertion of picture-changing operators
in the action~\cite{Saroja:1992vw}.
The resulting structure, however, is complicated and noncanonical
because one must give a prescription
to deform the closed string vertices
and prevent the collision of picture-changing operators.\footnote{
Open superstring field theory actions with explicit picture-changing
operators were considered in \cite{Witten:1986qs}.}

In this paper we construct a Lorentz-covariant string field theory
for the NS sector
of heterotic strings~\cite{Gross:1985fr}.
We work in the large Hilbert space
and use the nonpolynomial structure
of bosonic closed string field theory
to produce a cover of the moduli spaces of punctured spheres.
Gauge invariance further requires
elementary interactions associated with
the {\em boundary} of the missing regions of the moduli spaces
and with lower-dimensional subspaces of this boundary.
As in the case of WZW-like open superstring field theory,
no insertions of picture-changing operators are necessary.
Our work was partially motivated by the desire to have calculable
closed superstring field theories to investigate
conjectures concerning twisted tachyons in orbifold
backgrounds~\cite{Adams:2001sv}.

\section{Open superstring and closed bosonic string field theories}

In this section we review the structure of
WZW open superstring field theory
and bosonic closed string field theory,
focusing on the aspects necessary
for our construction of heterotic string field theory.
We begin with the open superstring field
theory~\cite{Berkovits:1995ab}.

A general off-shell open string field
configuration in the GSO-even NS sector corresponds to
a Grassmann {\it even} state
$|\Phi\rangle$ of ghost number zero
and picture number zero in the combined
conformal field theory (CFT)
of matter, ghosts, and superghosts.
This string field lives in the `large Hilbert space'
that contains the zero mode $\xi_0$ of the field $\xi$.
For the superghosts $\xi$, $\eta$, and $\phi$,
the assignments of ghost number ($G$) and picture number ($P$)
are as follows:
\be \label{ei2}
\xi:\quad G=-1 \,, ~P=1 \,, \qquad
\eta:\quad G=1 \,, ~P=-1 \,, \qquad
e^{q\phi}:\quad G=0 \,, ~P=q \,.
\ee
The string field theory action takes the following form:
\be \label{e0}
S= {1\over 2g^2} \, \Bigl\langle\Bigl\langle \,
(e^{-\Phi} \hskip1pt Q \hskip1pt e^{\Phi}) \,
(e^{-\Phi} \hskip1pt \eta_0 \hskip1pt e^\Phi)
- \int_0^1 dt \,
(e^{-t\Phi} \hskip1pt \partial_t \hskip1pt e^{t\Phi}) \,
\{ \, (e^{-t\Phi} \hskip1pt Q \hskip1pt e^{t\Phi}), \,
(e^{-t\Phi} \hskip1pt \eta_0 \hskip1pt e^{t\Phi}) \, \} \,
\Bigr\rangle\Bigr\rangle \,,
\ee
where $\{ A,\, B \} \equiv AB + BA$, $Q$ is the BRST operator,
and $\eta_0 = \oint dz \, \eta(z)$.
This action is defined
using Witten's star product \cite{Witten:1985cc}
by expanding all exponentials in formal Taylor series
preserving the order of all operators.
To cubic order one finds
\be \label{e17p}
S = {1\over 2g^2} \, \Bigl\langle\Bigl\langle
\,\,{1\over 2} \,
(Q \Phi) \, (\eta_0 \Phi) \,
+ {1\over 6} \,
(Q \Phi) \, \Bigl( \, \Phi \, (\eta_0 \Phi)
- (\eta_0 \Phi) \, \Phi \, \Bigr) \, \Bigr\rangle\Bigr\rangle
{}+ \mathcal{O}(\Phi^4) \,.
\ee
The full action is invariant under gauge transformations
with gauge parameters $\Lambda$ and $\Omega$:
\be \label{egtrs}
\delta e^{\Phi}
= (Q \Lambda) \, e^{\Phi} + e^{\Phi} \,
(\eta_0 \hskip1pt \Omega) \,.
\ee
The gauge invariance can be proven using, among others,
$\{ Q, \eta_0\} = 0$ and  $Q^2 = \eta_0^2 = 0$.
The equation of motion for the string field is
$\eta_0 \, (e^{-\Phi} \hskip1pt Q \hskip1pt e^{\Phi}) =0$,
which, to linearized order, reduces to
\begin{equation}
\label{linfeqn}
Q \hskip1pt \eta_0 \hskip1pt |\Phi\rangle=0\,.
\end{equation}
Using the $\Omega$ gauge invariance
one can choose the gauge $\xi_0 \hskip1pt |\Phi\rangle=0$,
and thus write
$|\Phi\rangle = \xi_0 \hskip1pt | \widehat{\Phi} \rangle$,
where $\eta_0 \hskip1pt |\widehat{\Phi} \rangle=0$.
Here $| \widehat{\Phi} \rangle$
is a state in the `small Hilbert space',
the space that does not include the zero mode of $\xi$.
In the gauge
$|\Phi\rangle = \xi_0 \hskip1pt | \widehat{\Phi} \rangle$,
the linearized equation of motion reduces to
$Q \hskip1pt | \widehat{\Phi} \rangle =0$.
This equation of motion is satisfied when the vertex operator
corresponding to the state $| \widehat{\Phi} \rangle$
takes the form of $c \hskip1pt e^{-\phi} V_M$, where
$V_M$ is  a matter primary  with dimension 1/2.
The corresponding state
$|\Phi \rangle = \xi_0 \hskip1pt | \widehat{\Phi} \rangle$
has ghost number and picture number zero, as expected.
In the GSO even sector $V_M$ is Grassmann odd.
Since $\xi_0$, $c$, and $e^{-\phi}$ are all Grassmann odd,
$|\Phi\rangle$ is Grassmann even.

\bigskip
Let us now turn to the second ingredient of the construction.
The bosonic closed string field theory action
\cite{Zwiebach:1992ie} is given by
\begin{equation}
\label{action}
S = -{2\over \alpha'} \, \Bigl(~ {1\over 2}
\langle \, \Psi,\,  Q \Psi \, \rangle
+ {1\over 3!} \, \kappa \,
\langle \, \Psi,\, [\, \Psi,\, \Psi \,] \, \rangle
+ {1\over 4!} \, \kappa^2 \,
\langle \, \Psi,\, [\, \Psi,\, \Psi,\, \Psi \,] \, \rangle
+ \, \ldots \,\Bigr) \,.
\end{equation}
The closed string field  $|\Psi\rangle$ has ghost number two
and satisfies the subsidiary conditions
$(b_0-\bar b_0) \hskip1pt |\Psi\rangle
= (L_0 - \bar L_0) \hskip1pt |\Psi\rangle =0$.
There is an infinite set of string products,
all of which are graded-{\em commutative}.
For example, the lowest product satisfies
$[ \, A,\, B \, ] = (-1)^{AB} \, [\, B,\, A \, ]$,
where string states in the exponent represent
their Grassmann property,
$0$ (mod 2) for Grassmann even states
and $1$ (mod 2) for Grassmann odd states.
The linear inner product is defined
by $\langle \, A \,, B \, \rangle
= \langle \, A \, | \hskip1pt c_0^- \hskip1pt | \, B \, \rangle$,
where $\langle  \, A \, |$
is the BPZ conjugate of $| \, A \, \rangle$,
and $c_0^- ={1\over 2} (c_0- \bar c_0)$.
Some important identities satisfied by the lowest product,
the inner product, and the BRST operator $Q$ are
\begin{eqnarray}
\label{lowerident}
\langle \, A,\, B \, \rangle
&=& (-1)^{(A+1) (B+1)} \, \langle \, B,\, A \, \rangle \,,
\nonumber\\
\langle \, QA,\, B \, \rangle
&=& (-1)^{A} \, \langle \, A,\, QB \, \rangle \,,\\
\langle \, [\, A,\, B \,] \,,\, C \, \rangle
&=& (-1)^{A+B} \, \langle \, A,\, [\, B,\, C \, ] \, \rangle\,.
\nonumber
\end{eqnarray}
The products, together with the inner product,
define fully graded-commutative multilinear forms:
\begin{equation}
\{ \, B_1,\, B_2,\, \ldots ,\, B_n \, \}
\equiv \langle \, B_1,\, [\, B_2,\, \ldots ,\, B_n \, ]\, \rangle\,.
\label{multilinear-forms}
\end{equation}
The fundamental relation satisfied by the closed string
products takes the form:
\begin{eqnarray}
\label{mainidentity}
0 &=& \hskip-5pt Q \, [\, B_1,\, \ldots ,\, B_n \,]
+ \sum_{i=1}^n (-1)^{(B_1+ \ldots + B_{i-1})}
[\, B_1,\, \ldots, \, Q B_i,\, \ldots ,\, B_n \,] \nonumber\\
&&\hskip-5pt + \sum_{\{i_l,\, j_k\}}
\sigma(i_l,\, j_k) \,
[\, B_{i_1},\, \ldots,\, B_{i_l},\,
[ \, B_{j_1},\, \ldots ,\,  B_{j_k} \, ] \,] \,.
\end{eqnarray}
The sum in the second line runs
over all different splittings of the set
$\{ 1, 2, \ldots, n\}$
into a first group $\{ i_1, \ldots , i_l\}$
and a second group $\{ j_1, \ldots, j_k\}$,
where $l \ge 1$ and $k \ge 2$.
Two splittings are the same if the corresponding first groups
contain the same set of integers regardless of their order.
The sign factor $\sigma(i_l,\, j_k)$
is the sign picked up when one rearranges the sequence
$\{ \, Q,\, B_1,\, B_2,\, \ldots \,,\, B_n\}$
into the sequence $\{ \, B_{i_1},\, \ldots,\, B_{i_l},\, Q,\,
B_{j_1},\, \ldots,\, B_{j_k} \, \}$ taking into account
the Grassmann property of the various objects.
For $n=2$, the identity is
\begin{equation}
0= Q \, [\, B_1,\, B_2 \,] + [\, QB_1,\, B_2 \,]
+ (-1)^{B_1} \, [\, B_1, \, QB_2 \,]\,.
\label{n=2-identity}
\end{equation}
This is the derivation property of $Q$.
One might have expected a minus sign in front of the first term
on the right-hand side, but the sign is absent
because the string product carries an insertion of $b_0 - \bar b_0$.
For $n=3$, the identity (\ref{mainidentity}) is
\begin{eqnarray}
0 &=& \hskip-5pt Q \, [\, B_1,\, B_2,\, B_3 \,]
+ [\, QB_1,\, B_2,\, B_3 \,]
+ (-1)^{B_1} \, [\, B_1,\, QB_2,\, B_3 \,]
+ (-1)^{B_1+B_2} \, [\, B_1,\, B_2,\, QB_3 \,]
\nonumber\\[1ex] &&\hskip-8pt
{}+ (-1)^{B_1} \, [\, B_1,\, [\, B_2,\, B_3 \,] \,]
+ (-1)^{B_2 (1+ B_1)} \, [\, B_2,\, [\, B_1,\, B_3 \,] \,]
+ (-1)^{B_3 (1+B_1+B_2)} \, [\, B_3,\, [\, B_1,\, B_2 \,] \,] \,.
\nonumber \\
\end{eqnarray}
Because of insertions of $b$ ghosts, $Q$ is no longer a derivation.
The violation of the derivation property of $Q$
is related to the violation of the Jacobi identity.

The ghost number of a string product
is not the sum of the ghost number of each state.
For a product with $n$ input states we have
\begin{equation}
\label{ghnumb}
G( \, [ \, B_1,\, B_2,\, \ldots,\, B_n \, ] \, )
= -2(n-2) - 1 + \sum_{i=1}^n G_i \,,
\end{equation}
where $G_i$ is the ghost number of $B_i$.

\sectiono{The construction of Heterotic String Field Theory}
\label{section3}

The string field theory that we will construct can be
formulated for any consistent background of heterotic string theory,
namely, for any conformal field theory
with the following structure
of holomorphic and antiholomorphic sectors.
The holomorphic sector is comprised of
an $N=1$ superconformal matter theory with central charge $c=15$,
reparameterization ghosts $(b,c)$ with central charge $c=-26$,
and superghosts $(\xi, \eta, \phi)$ with $c=11$.
The antiholomorphic sector is not supersymmetric and consists of
a matter theory with $\bar{c}=26$
and reparametrization ghosts $(\bar b, \bar c)$
with $\bar{c}=-26$.
The full theory has an NS sector and a Ramond (R) sector,
depending on the boundary conditions of fermions
on the supersymmetric side.
In this paper we focus on the NS sector and assume
that this sector has been truncated to the GSO even states.
Presumably, a formulation that describes both
GSO even and odd states can be obtained with small modifications,
as discussed in detail in~\cite{Berkovits:2000zj}.

\bigskip

The open superstring field was defined to be Grassmann even
and had both ghost number and picture number zero.
In bosonic open string field theory the string field
is Grassmann odd,  has ghost number one,
and carries no picture.
By tensoring the two string fields,
the resulting closed string field $V$
should be Grassmann odd and have ghost number one
and picture number zero:
\begin{equation}
V \,:\, \hbox{Grassmann odd},  ~~G=1, ~~ P=0\,.
\end{equation}
We work in the large state space of the superconformal sector.
Physical vertex operators
corresponding to $| V \rangle$
in the gauge $\xi_0 | V \rangle = 0$
take the form of
$\mathcal{V}= \xi  \, c \bar c \, V_M \, e^{-\phi}$.
Here $V_M$ is a Grassmann odd matter primary operator
with dimensions $(1/2, \, 1)$. Since $\mathcal{V}$
is Grassmann odd, so is the string field.
The normalization of correlators in the
full CFT for a flat spacetime background is given by
\be \label{eb2}
\langle\langle \, \xi(w_1)~ e^{-2\phi(w_2)} \,
c(z_1) \bar{c}(\bar{z}_1) \, c(z_2) \bar{c}(\bar{z}_2) \,
c(z_3) \bar{c}(\bar{z}_3) \, e^{i p \cdot X (z,\bar{z})} \,
\rangle\rangle
= 2 (2 \pi)^D \delta^{(D)} (p)
|z_1-z_2|^2 |z_1-z_3|^2 |z_2-z_3|^2 \,,
\ee
where $D=10$ is the dimension of spacetime.
Nonvanishing
correlators require total ghost number five
and total picture number minus one.
Since we will write the action
using the inner product $\langle \, \cdot\,,\,\cdot \, \rangle$
which contains an insertion of $c^-_0$,
each term in the action must have
ghost number four and picture number minus one.

In the full CFT we have a BRST operator $Q$ that is
nilpotent and, as usual, satisfies the relations $\{Q, b(z)\} = T(z)$
and $\{Q, \bar b(\bar z)\} = \bar T(\bar z)$, with $T, \bar T$
the total stress tensor of the conformal field theory.
Since we are dealing with a closed string theory,
the following subsidiary conditions apply:
\begin{equation}
(b_0-\bar b_0) \hskip1pt |V\rangle =
(L_0 - \bar L_0) \hskip1pt |V\rangle =0 \,.
\label{subsidiary-conditions}
\end{equation}
The inner product is
$\langle \, A \,, B \, \rangle
= \langle \, A \, | \hskip1pt c_0^- \hskip1pt | \, B \, \rangle$,
formally the same one we had for bosonic closed strings.
The string products are obtained by integrating forms
over certain subsets of the ordinary moduli spaces of
punctured Riemann spheres.
The forms have no superghost insertions, while
insertions of $b$ antighosts are present in exactly the same way
as in bosonic closed string field theory.
This implies that the string products in heterotic strings
satisfy the identity~\refb{mainidentity}.
The identities in \refb{lowerident} also hold.
For simplicity, we denote
the zero mode of the field $\eta$ by $\eta$ itself.
Then we have
\begin{equation}
\label{etabpz}
\langle \, \eta A,\, B \, \rangle
= (-1)^{A} \, \langle \, A,\, \eta B \, \rangle \,.
\end{equation}
Since the string products contain no
superghost insertions, $\eta$ is a derivation:
\begin{equation}
0 = \eta\, [\, B_1,\, \ldots,\, B_n \,]
+ \sum_{i=1}^n (-1)^{(B_1+ \ldots + B_{i-1})} \,
[\, B_1,\, \ldots,\, \eta B_i,\, \ldots,\, B_n \,] \,.
\label{eta-derivation}
\end{equation}

\subsection{Quadratic and cubic terms in the action}

The kinetic term for the heterotic string field theory action
is expected to give the linearized field equation
$Q \hskip1pt \eta \hskip1pt | \, V \, \rangle =0 \,$,
which in the gauge $\xi_0 \hskip1pt | \, V \, \rangle =0$
reduces to the conventional BRST cohomology problem
in the small Hilbert space.
We expand the action
in powers of the gravitational constant $\kappa$:
\begin{equation}
S = \frac{2}{\alpha'} \sum_{n=2}^{\infty} \kappa^{n-2} S_n \,,
\end{equation}
and we write $S_2$ as
\begin{equation}
S_2= {1\over 2} \,\langle \, \eta V,\, QV \, \rangle\,.
\end{equation}
As required, the total ghost number and total picture number for
the operators in the above inner product are four and minus one,
respectively.
Using \refb{lowerident} and \refb{etabpz}
the variation of $S_2$ is given by
\begin{equation}
\label{varkin}
\delta S_2
= \langle \, \delta V,\, Q \hskip1pt \eta V \, \rangle\,
= \langle \, Q \hskip1pt \eta V,\, \delta V\, \rangle\,,
\end{equation}
which yields the correct linearized equation of motion.
The kinetic term is invariant under the transformations
\begin{equation}
\delta_\Lambda^{(0)} V = Q\Lambda\,, \qquad
\delta_\Omega^{(0)} \hskip1pt V = \eta \hskip1pt \Omega\,.
\end{equation}

The cubic interaction requires the lowest string product.
Since $V$ is Grassmann odd
and the string products are graded-commutative,
the product $[ \, V,\, V \, ]$ vanishes.
On the other hand,
$[\, V,\, QV \,]$ does not vanish, and it has ghost number two,
as can be seen from \refb{ghnumb}.
We write $S_3$ as
\begin{equation}
S_3 = {1\over 3!} \, \langle \, \eta V,\,
[\, V,\, QV \,] \, \rangle \,,
\end{equation}
or, using the multilinear form, as
\begin{equation}
S_3 = {1\over 3!} \, \{ \, \eta V,\, V,\, QV \, \} \,.
\end{equation}
Since the sum of pictures of the operators involved
must be minus one, there is just one factor of $\eta$.
This will be the case for all terms in the action.
It is instructive to see how the above cubic term reduces
to the expected correlator for physical states.
First we note that
$\{\, A,\, B,\, C \,\}
= \langle \langle \, A(\infty) \, B(0) \, C(1) \,
\rangle\rangle \,$ for physical states,
where $\langle\langle \, \ldots \, \rangle \rangle$
denotes correlator. Writing
$| V \rangle = \xi_0 \hskip1pt | \widehat V \rangle$,
where $Q \hskip1pt | \widehat V \rangle=0$
and $| \widehat V \rangle$ has picture minus one,
we see that
\begin{equation}
S_3 = \frac{1}{3!} \, \langle\langle \,
\{ \hskip1pt \eta_0 \hskip1pt, V (\infty) \hskip1pt \} \,\,
V(0) \,\,
\{ \hskip1pt Q, V (1) \hskip1pt \} \, \rangle\rangle
= \frac{1}{3!} \,
\langle\langle \, \widehat V (\infty) \,\,
\xi \widehat V (0) \,\,
\oint \frac{dz}{2 \pi i z} \, X(z) \, \widehat V (1) \,
\rangle\rangle\,,
\end{equation}
where $X(z) = \{ \hskip1pt Q,\, \xi (z) \hskip1pt \}$
is the picture-changing operator
and the contour of the integral encircles $z=1$ counterclockwise.
The expression on the right-hand side,
using \refb{eb2}, reduces to the small Hilbert space correlator
of $\widehat V (\infty)$, $\widehat V (0)$,
and $\lim_{z \to 1} X(z) \hskip1pt \widehat V (1)$,
the last of which is the vertex operator in the zero picture.
This is the expected result.

To see how the gauge transformations are modified by the
addition of the cubic interaction
we first consider the general variation of $S_3$.
Using the following formulas
\begin{eqnarray}
0 &=& \{ \, Q B_1,\, B_2,\, B_3 \, \}
+ (-1)^{B_1} \, \{ \, B_1,\, Q B_2,\, B_3 \, \}
+ (-1)^{B_1 + B_2} \, \{ \, B_1,\, B_2,\, Q B_3 \, \} \,,
\nonumber \\
0 &=& \{ \, \eta \,B_1,\, B_2,\, B_3 \, \}
+ (-1)^{B_1} \, \{ \, B_1,\, \eta\, B_2,\, B_3 \,\, \}
+ (-1)^{B_1 + B_2} \, \{ \, B_1,\, B_2,\, \eta\, B_3 \, \} \,,
\end{eqnarray}
which follow from (\ref{lowerident}), (\ref{multilinear-forms}),
(\ref{n=2-identity}), (\ref{etabpz}), and (\ref{eta-derivation}),
a short calculation gives
\begin{equation}
\delta S_3 = {1\over 2} \, \{ \, \delta V,\,
\eta V,\, QV \, \}
= {1\over 2} \, \langle \, \delta V,\,
[\, \eta V,\, QV \,] \, \rangle \,.
\end{equation}
It follows that the gauge variation $\delta_\Lambda^{(0)} S_3$
is given by
\begin{equation}
\delta_\Lambda^{(0)} S_3 = {1\over 2} \, \langle \, Q\Lambda,\,
[\, \eta V,\, QV \,] \, \rangle \,.
\end{equation}
Using $[\, \eta V,\, QV \,]
= - Q \, [\, \eta V,\, V \,]
- [\, Q \hskip1pt \eta V,\, V \,] \,$, we obtain
\begin{equation}
\delta_\Lambda^{(0)} S_3
= -{1\over 2} \, \langle \, Q\Lambda,\,
[\, Q \hskip1pt \eta V,\, V \,] \, \rangle
= -{1\over 2} \, \langle \, Q \hskip1pt \eta V,\,
[\, V,\, Q\Lambda \,] \, \rangle \,.
\end{equation}
Making use of \refb{varkin}
we deduce that the above variation of the cubic term
can be cancelled against a variation of the quadratic term
by modifying the gauge transformation:
\begin{equation}
\delta_\Lambda V = Q\Lambda
+ {\kappa\over 2} \, [\, V,\, Q\Lambda \,]
+ \mathcal{O} (\kappa^2) \,.
\end{equation}
The gauge transformation generated by $\eta$ works out
in a similar way:
\begin{equation}
\delta_\Omega^{(0)} S_3 =
{1\over 2} \, \langle \, \eta \hskip1pt \Omega,\,
[\, \eta V,\, QV \,] \, \rangle \,.
\end{equation}
Using $[\, \eta V,\, QV \,]
= - \eta \, [\, V,\, QV \,]
- [\, V,\, Q \hskip1pt \eta V \,] \,$,
we find
\begin{equation}
\delta_\Omega^{(0)} S_3
= -{1\over 2} \, \langle \, \eta \hskip1pt \Omega,\,
[\, V,\, Q \hskip1pt \eta V \,] \, \rangle
= -{1\over 2} \, \langle \, Q \hskip1pt \eta V,\,
[\, \eta \hskip1pt \Omega,\, V \,] \, \rangle \,.
\end{equation}
This variation is also cancelled
by modifying the gauge transformation:
\begin{equation}
\delta_\Omega \hskip1pt V
= \eta\hskip1pt\Omega + {\kappa\over 2} \,
[\, \eta\hskip1pt\Omega,\, V \,] + \mathcal{O} (\kappa^2)\,.
\end{equation}
To cubic order the action is
\begin{equation}
\label{cubactionepiojdh}
S = \frac{2}{\alpha'} \biggl[ \,
{1\over 2} \, \langle \, \eta V,\, QV \, \rangle
+ {\kappa\over 3!} \, \langle \, \eta V,\,
[\, V,\, QV \,] \, \rangle \, \biggr] + \mathcal{O} (\kappa^2)\,,
\end{equation}
and the equation of motion takes the form:
\begin{equation}
Q \hskip1pt \eta V + {\kappa\over 2} \, [\, \eta V,\, QV \,]
+ \mathcal{O}(\kappa^2) = 0 \,.
\end{equation}
We have computed
the on-shell scattering amplitude of three gravitons
using the action \refb{cubactionepiojdh}.
We have confirmed that
$\kappa$ is the gravitational constant
by comparing our result with the properly normalized amplitude
in equation (12.4.14) of~\cite{polch}.

\subsection{Higher-order terms}

In this subsection we first make a few remarks
on the general structure of higher-order terms
in the string field theory action.
We then list our results for the quartic and quintic terms,
together with the corresponding gauge transformations.
Finally, we determine certain classes of terms in the action
to all orders.

\medskip

We have already learned that each term in the action
has one factor of $\eta$.
Using \refb{lowerident}, (\ref{multilinear-forms}),
the graded-commutativity of the multilinear forms,
and possibly \refb{mainidentity},
we can always write any interaction as an inner product
of $\eta V$ with a string field
build with string products of $V$'s and $QV$'s:
\begin{equation}
\langle \, \eta V,\, [\, \ldots \, [ \, \ldots \,
[\, \ldots \, ] \, \ldots \,
[\, \ldots \, ] \, \ldots \, ] \, ] \, \rangle \,.
\end{equation}
Consider how many times $QV$ can appear
in the terms of $\mathcal{O}(V^{N+1})$.
The minimum number of products is one, in which case the product
has $N$ entries and the term is
\begin{equation}
\label{topterm}
\langle \, \eta V ,\,
[\, V,\, (QV)^{N-1} \,] \, \rangle\,
\equiv \langle \, \eta V,\, [\, V,\, \underbrace{QV,\, QV,\,
\ldots,\, QV}_{N-1} \,\,] \, \rangle \,.
\end{equation}
Here we introduced the shorthand notation $(QV)^{N-1}$
for a collection of $N-1$ entries of $QV$.
There cannot be more than one $V$
since any product with multiple $V$'s vanishes
by graded-commutativity.
The $(N-1)$ $QV$'s are in fact necessary
for the ghost number to work.
One can easily see from \refb{ghnumb}
that the string product has ghost number two
when there are $(N-1)$ $QV$'s.
The maximum possible number of products
in the terms of $\mathcal{O}(V^{N+1})$
is $N-1$.
In this case, the interaction term takes the following form:
\begin{equation}
\label{lowterm}
\langle \, \eta V,\,
\underbrace{\,[\,V,\, [\, V,\,
[\, V,\, \ldots \, [\, V,}_{N-1} \, QV \,] \,] \,] \ldots \,
] \, \rangle \,.
\end{equation}
To investigate the general constraint from ghost number,
consider a product with $N$ inputs:
\begin{equation}
[\, B_1,\, B_2,\, \ldots,\, B_N \,] \,.
\end{equation}
We can introduce additional products by writing
additional pairs of brackets inside the outer brackets.
If we add one pair of brackets, we have two products,
one with $N_1$ inputs and one with $N_2$ inputs,
where $N_1 + N_2 = N+1$.
In fact, each time we add a pair of brackets
the number of inputs increases by one unit.
For a term built with $k$ products with input numbers
$N_1,\, N_2,\, N_3,\, \ldots,\, N_k \,$,
the total number of inputs is $N+k-1$.
Using \refb{ghnumb} the ghost number of the term in question is
\begin{equation}
G= \sum_{j=1}^k ( -2N_j + 3)+\sum_{i=1}^N  G_i
= -2 (N+ k-1) + 3 k+\sum_{i=1}^N  G_i
= -2N + k + 2 + \sum_{i=1}^N  G_i \,.
\end{equation}
Since we need $G=2$, we learn that
$\sum_{i=1}^N  G_i =  2N - k $.
With $V$'s and $QV$'s the only possible inputs,
we must use $k$ copies of $V$ and $N-k$ copies of $QV$.
In summary, {\em
$S_{N+1}$ consists of terms of the form
$\langle \, \eta V,\, R_k \, \rangle \,$,
where $1\leq k\leq N-1$, and $R_k$ is a state built
using $k$ string products,
$k$ copies of $V$, and $N-k$ copies of $QV$}.
{\em Each product has exactly one entry that is~$V$.}

\medskip
Let us look at the quartic and quintic terms explicitly.
The above analysis tells us that there are two possible terms
in the quartic part $S_4$ of the action:
the first uses a single product with three inputs,
and the second uses two products each of which has two inputs.
We have found by computation that gauge invariance
uniquely determines the coefficients for these terms to be
\begin{equation}
\label{quarticterms}
S_4 = \frac{1}{4!} \, \Bigl(
\langle \, \eta V,\, [\, V,\, QV,\, QV \,] \, \rangle\,
+ \langle \, \eta V,\, [\, V,\, [\, V,\, QV \,] \,] \,
\rangle \, \Bigr) \,.
\end{equation}
Geometrically, the first term is a correlator integrated
over the subspace $\mathcal{V}_{0,4}$
of the moduli space $\mathcal{M}_{0,4}$ of four-punctured spheres.
The subspace $\mathcal{V}_{0,4}$ is determined canonically in closed
string field theory using minimal area metrics,
and it has an explicit description in terms of polyhedra
associated with Jenkins-Strebel quadratic differentials.
In the computation of string amplitudes, this term is necessary
to produce a cover of  the moduli space $\mathcal{M}_{0,4}$.
The second term in \refb{quarticterms} is a correlator
integrated over a part of the boundary of $\mathcal{V}_{0,4}$.
Although it is required by gauge invariance,
it does not contribute to on-shell scattering amplitudes
when the vertex operators take the form of
$\xi \, c \bar{c} \, e^{-\phi}$ times a primary operator
in the matter sector.

With the inclusion of $S_4$,
the gauge transformations of the string field acquire terms
quadratic in $V$. One finds
\begin{eqnarray}
\delta_\Lambda V
&=& Q\Lambda + {\kappa\over 2} \, [\, V,\, Q\Lambda \,]
+ \kappa^2 \, \Bigl( \, {1\over 6} \, [\, V,\, QV,\, Q\Lambda \,]
+ {1\over 12} \, [\, V,\, [\, V,\, Q \Lambda \,] \,] \,\Bigr)
+ O(\kappa^3) \,,
\\
\delta_\Omega \hskip1pt V
&=& \eta\hskip1pt\Omega
+ {\kappa\over 2} \, [\, \eta\hskip1pt\Omega\hskip1pt,\, V \,]
+ \kappa^2 \, \Bigl(\,
{1\over 3} \, [\, \eta\hskip1pt\Omega\hskip1pt,\, QV,\, V \,]
+ {1\over 12} \, [\, [\, \eta\hskip1pt\Omega\hskip1pt,\, V \,] \,,
V \,] \, \Bigr) + O(\kappa^3) \,.
\end{eqnarray}
The gauge transformations are not uniquely determined
since we can redefine the gauge parameters.
We have fixed this ambiguity by the requirement
that $\Lambda$ always appear as $Q\Lambda$
and $\Omega$ always appear as $\eta\hskip1pt\Omega$.
The correction to the equation of motion follows from
the general variation of the quartic terms:
\begin{eqnarray}
&& \delta \, \bigl( \,
\langle \, \eta V,\,
[\, V,\, QV,\, QV \,] \, \rangle\,
+ \langle \, \eta V,\, [\, V,\, [\, V,\, QV \,] \,] \,
\rangle \, \bigr)
\nonumber \\
&=& 4 \, \langle \, \delta V,\,
[\, QV,\, QV,\, \eta V \,] \, \rangle
-2 \, \langle \, \delta V,\,
[\, V,\, [\, QV,\, \eta V \,] \,] \, \rangle
\nonumber \\ && \hskip-13pt {}
+4 \, \langle \, \delta V,\,
[\, \eta V,\, [\, V,\, QV \,] \,] \, \rangle
+2 \, \langle \, \delta V,\,
[\, V,\, [\, V,\, Q \hskip1pt \eta V \,] \,] \, \rangle \,.
\end{eqnarray}

We have also determined
the quintic part $S_5$ of the action completely.
This time there are four terms:
one term with one product, two inequivalent terms with
two products, and one term with three products.
Their coefficients are again uniquely determined
by gauge invariance. We found
\begin{eqnarray}
\label{quinticac}
S_5 &=& {1\over 5!} \, \Bigl(
\, \langle \, \eta V,\, [\, V,\, QV,\, QV,\, QV \,] \, \rangle\,
+ \langle \, \eta V,\, [\, V,\, [\, V,\, QV,\, QV \,] \,] \, \rangle
\nonumber \\[0.5ex]
&&\quad {}+ 3 \, \langle \, \eta V,\, [\, V,\, QV,\,
[\, V,\, QV \,] \,] \,\rangle
+ \langle \, \eta V,\, [\, V,\, [\, V,\,
[\, V,\, QV \,] \,] \,] \,\rangle \, \Bigr) \,.
\end{eqnarray}
The gauge transformations $\delta_\Lambda V$
and $\delta_\Omega \hskip1pt V$ acquire the following extra terms:
\begin{eqnarray}
\delta_\Lambda^{(3)} V &=& {1\over 4!} \,
\Bigl( \, [\, V,\, QV,\, QV,\, Q\Lambda \,]
+ [\, V,\, QV,\, [\, V,\, Q\Lambda \,] \,]
+ [\, [\, V,\, QV \,] \,,\, V,\, Q\Lambda \,] \, \Bigr) \,,
\\[1ex]
\delta_\Omega^{(3)} \hskip1pt V &=& {1\over 8} \,
[\, \eta\hskip1pt\Omega\hskip1pt,\, QV,\, QV,\, V \,]
+ {1\over 12} \, [\, [\, \eta\hskip1pt\Omega\hskip1pt,\,
QV,\, V \,] \,, V \,]
\nonumber \\
[0.5ex]
&& \hskip-8pt + {1\over 8} \, [\, \eta\hskip1pt\Omega\hskip1pt,\,
[\, QV,\, V \,] \,,\, V, \,]
+ {1\over 24} \, [\, [\, \eta\hskip1pt\Omega\hskip1pt,
V \,] \,, QV,\, V \,] \,,
\end{eqnarray}
where $\delta_\Lambda^{(m)} V$
and $\delta_\Omega^{(m)} \hskip1pt V$
are defined by
\begin{equation}
\delta_\Lambda V
= \sum_{m=0}^{\infty} \kappa^m \, \delta_\Lambda^{(m)} V \,,
\quad
\delta_\Omega \hskip1pt V
= \sum_{m=0}^{\infty} \kappa^m \,
\delta_\Omega^{(m)} \hskip1pt V \,.
\end{equation}

\medskip

While we have not determined all terms
in the string field theory action
and gauge transformations, we have completely determined
the terms in the action that are built with one string product
and with two string products.
We have also derived the terms in the gauge transformations
that are built with one string product.
For the action we find
\begin{equation}
S_n = {1\over n!} \, \Bigl( \, \langle \, \eta V,\,
[\, V,\, (QV)^{n-2} \,] \, \rangle
+ \sum_{m=0}^{n-4} { \, n-2 \, \choose m } \,
\langle \, \eta V,\, [\, V,\, (QV)^m,\,
[\, V,\, (QV)^{n-m-3} \,] \,] \, \rangle \, \Bigr)
+ \mathcal{O}(Q^{n-4}) \,,
\label{S_ndetailedexpansion}
\end{equation}
and for the gauge transformations we find
\begin{eqnarray}
\delta_\Lambda^{(m)} V
&=& {1\over (m+1)!} \, [\, V,\, (QV)^{m-1},\, Q \Lambda \,]
+ \mathcal{O}(Q^{m-1}) \,,
\label{general-Lambda-gauge-transformation} \\
\delta_\Omega^{(m)} \hskip1pt V
&=& {m\over (m+1)!} \,
[\, \eta \hskip1pt \Omega,\, (QV)^{m-1},\, V \,]
+ \mathcal{O}(Q^{m-2}) \,.
\label{general-Omega-gauge-transformation}
\end{eqnarray}
We present the derivation of
(\ref{S_ndetailedexpansion}),
(\ref{general-Lambda-gauge-transformation}),
and (\ref{general-Omega-gauge-transformation}) in the Appendix.
We found that gauge invariance under either
$\delta_\Lambda^{(m)} V$ or $\delta_\Omega^{(m)} \hskip1pt V$ alone
determines the coefficients
in (\ref{S_ndetailedexpansion}) uniquely
and gives the same set of coefficients.
We regard this as evidence that
a unique gauge-invariant action exists to all orders.

\medskip
Let us conclude the section by making some comments
on the geometrical meaning
of the various interactions in the action.
In bosonic closed string field theory
the moduli space $\mathcal{M}_{0,n}$ of $n$-punctured spheres
is covered by the contributions
of an elementary vertex $\mathcal{V}_{0,n}$ and Feynman diagrams
built with lower-order vertices and propagators.
One writes
\begin{equation}
\label{setfullm}
\mathcal{M}_{0,n}
= \mathcal{V}_{0,n} \cup \mathcal{R}_{\,0,n}^{(1)}
\cup \mathcal{R}_{\,0,n}^{(2)}\cup \ldots
\cup \mathcal{R}_{\,0,n}^{(n-3)}\,,
\end{equation}
where $\mathcal{R}_{\,0,n}^{(I)}$ denotes
the region of moduli space
obtained by Feynman graphs with $I$ propagators.
As indicated, the maximum number of propagators is $n-3$.
The Feynman graphs in $\mathcal{R}_{\,0,n}^{(n-3)}$ are
built using only the three string vertex,
and all the moduli arise from the propagators,
each of which carries two moduli, length and twist angle.
In bosonic closed string field theory,
the $n$-th order contribution to the action is
simply a correlator over $\mathcal{V}_{0,n}$.
This is all that is needed for gauge invariance.
In heterotic theory elementary vertices arise from
{\em all} components of \refb{setfullm}:
in addition to the correlator integrated over $\mathcal{V}_{0,n}$,
there are correlators integrated over subspaces
obtained by collapsing {\em all} propagators
of $\mathcal{R}_{\,0,n}^{(I)}$.
Each time we collapse a propagator we lose the length parameter,
but the twist-angle parameter survives.
For $\mathcal{R}_{\,0,n}^{(1)}$ we collapse the single propagator
and we obtain a subset of $\mathcal{M}_{0,n}$ of codimension one.
For $\mathcal{R}_{\,0,n}^{(2)}$ we collapse the two propagators
and we obtain a subset of $\mathcal{M}_{0,n}$ of codimension two.
For $\mathcal{R}_{\,0,n}^{(n-3)}$ we collapse the $n-3$ propagators
and we obtain a subset of $\mathcal{M}_{0,n}$ of codimension $n-3$,
namely, a subspace of dimension $n-3$,
which is half of the dimension of $\mathcal{M}_{0,n}$.
As a result, the dimensionalities
of the subspaces of $\mathcal{M}_{0,n}$
that define the vertices range from the top dimension
to half of that dimension.
The highest and lowest dimensional terms correspond,
in the case of $\mathcal{M}_{0,N+1}$,
to \refb{topterm} and \refb{lowterm}, respectively.
If we attempted to build
vertices associated with such lower dimensional subspaces
in bosonic string field theory,
they would vanish for string fields of ghost number two,
which is the ghost number of the classical string field.

For four-punctured spheres
we have $\mathcal{M}_{0,4}
= \mathcal{V}_{0,4} \cup \mathcal{R}_{\,0,4}^{(1)}$.
The two components correspond
to the two terms in \refb{quarticterms}.
For quintic terms we have $\mathcal{M}_{0,5}
= \mathcal{V}_{0,5} \cup \mathcal{R}_{\,0,5}^{(1)}
\cup \mathcal{R}_{\,0,5}^{(2)}$.
The first term in \refb{quinticac} corresponds
to an integral over $\mathcal{V}_{0,5}$,
the next two terms are integrals over $\mathcal{R}_{\,0,5}^{(1)}$
with the propagator collapsed,
and the last term is an integral over $\mathcal{R}_{\,0,5}^{(2)}$
with both propagators collapsed.
At each order the general set of vertices
that we have presented in (\ref{S_ndetailedexpansion})
correspond to the top subspace $\mathcal{V}_{0,n}$
and $\mathcal{R}_{\,0,n}^{(1)}$ with one propagator collapsed.
Note, however, that not all Feynman graphs produce
elementary interactions because, as we discussed earlier,
each string product must have one entry of $V$
for the ghost number to work.
If we define an internal vertex to be one all of whose legs
are not external legs of the Feynman graph,
the rule can be briefly stated as follows:
graphs with internal vertices
do not generate elementary interactions.

A canonical construction
of the subspaces $\mathcal{V}_{0,n}$ is obtained
in terms of restricted $n$-faced polyhedra:
polyhedra in which each face has perimeter $2\pi$
and  all nontrivial closed paths
along the edges are larger than or equal to $2\pi$.
The punctured sphere associated with a polyhedron
is built by attaching a semi-infinite cylinder
to each face of the polyhedron.
The surfaces obtained from $\mathcal{R}_{\,0,n}^{(I)}$
by collapsing all $I$ propagators
are in fact restricted polyhedra
in which $I$ nontrivial, non-intersecting closed paths
along the edges have length exactly equal to $2\pi$.
This gives a unified
description of all the moduli spaces
associated with heterotic string vertices
in terms of $\mathcal{V}_{0,n}$ and its subspaces.

\sectiono{Open questions and discussion}

\noindent
The most important remaining task is finding
full closed-form expressions for the string field theory action
and the gauge transformations.
Our partial results should give useful clues in this search.
The WZW open superstring field theory uses
exponentials of string fields.
This is natural because the string field is Grassmann even
and the open string star product is associative.
With the rich structure of homotopy Lie-algebra products
that appears in closed string field theory,
there may exist useful generalizations
of the exponential function.
Since the number of $QV$'s in the action increases
with the order of the interaction and $QV$ is Grassmann even,
some special function built with $QV$ may play a role
in the construction.
The two gauge transformations
$\delta_\Lambda V$ and $\delta_\Omega \hskip1pt V$
in heterotic string field theory
take rather different forms.
Each one separately appears
to determine the action completely.
It remains to show
the existence of a unique action which is invariant
under both gauge transformations to all orders.

Our construction is reminiscent of `heterosis',
the process of combining a right-moving superstring
with a left-moving bosonic string~\cite{Gross:1985fr}.
To see this, we recall that
one may be able to view bosonic closed string field theory
as a kind of tensor product
of two bosonic open string field theories~\cite{Gaberdiel:1997ia}.
In our case we combine an open superstring field theory
and an open bosonic string field theory,
the result being a closed string field theory
with a new algebraic structure.
While the algebraic structures
of both open and closed bosonic string field theories
were naturally understood in terms of master actions
that satisfy the Batalin-Vilkovisky master equation,
such understanding is still lacking
for the WZW open superstring theory
and, of course, for the present heterotic string field theory.

It would also be of interest to write the NS-NS sector
of a type II superstring field theory.
An intriguing generalization of the WZW open superstring theory
was discussed in~\cite{Berkovits:1998bt},
where the large Hilbert space was used
for both holomorphic and antiholomorphic sectors,
and a cubic covariantized light-cone theory was proposed.
For a Lorentz covariant theory higher-point interactions
of the form (\ref{multilinear-forms}) are necessary
to cover the moduli spaces of punctured spheres.
If we try to construct such a term with $N$ string fields
of ghost number and picture number zero
and operators $\eta$, $\bar{\eta}$, and $Q$
which preserve the conditions (\ref{subsidiary-conditions}),
the constraints from ghost and picture numbers
require $2N-4$ insertions of BRST operators.
Since this number is greater than the number of string fields
for large $N$, and we cannot have more than one $Q$
acting on a string field due to nilpotency,
our construction does not seem to extend
to the type II case in a straightforward way.

Despite its nonpolynomiality, the WZW open superstring field theory
is not significantly harder to use
than the cubic open bosonic string field theory.
This is because the higher point functions carry no moduli,
and no insertions of picture-changing operators are necessary.
We expect that the heterotic string field theory proposed here
will not be significantly harder to use
than bosonic closed string field theory.
We believe that computations of tachyon potentials
and investigations of spacetime background changes
are now feasible using heterotic string field theory.

\medskip
\noindent{\bf Acknowledgements}\\
\noindent
We would like to thank
Ashoke~Sen
for useful discussions.
This work was supported in part by
the DOE grant DF-FC02-94ER40818.

\medskip

\appendix

\sectiono{Constraints from gauge invariance} \label{a1}

We expand the gauge transformations of
$S$ in powers of $\kappa$:
\begin{equation}
\delta_\Lambda S
= \frac{2}{\alpha'}
\sum_{n=0}^{\infty} \kappa^n \, (\delta_\Lambda S)_n \,,
\quad
\delta_\Omega \hskip1pt S
= \frac{2}{\alpha'}
\sum_{n=0}^{\infty} \kappa^n \, (\delta_\Omega \hskip1pt S)_n \,.
\end{equation}
We can further classify the terms in $S_n$,
$(\delta_\Lambda S)_n$, and $(\delta_\Omega \hskip1pt S)_n$
by the number of BRST operators.
The maximum number of $Q$'s in $S_n$ is $n-2$ for $n \ge 3$.
Terms with $(n-2)$ $Q$'s
and terms with $(n-3)$ $Q$'s can be written as
\begin{equation}
S_n = f_n \, \langle \, \eta V,\,
[\, V,\, (QV)^{n-2} \,] \, \rangle
+ \sum_{m=0}^{n-4} f_{n, \, m} \,
\langle \, \eta V,\, [\, V,\, (QV)^m,\,
[\, V,\, (QV)^{n-m-3} \,] \,] \, \rangle
+ \mathcal{O}(Q^{n-4}) \,,
\label{S_n-expansion}
\end{equation}
for $n \ge 3$,
where $f_n$ and $f_{n, \, m}$ are coefficients
to be determined by gauge invariance.
The maximum number of $Q$'s in $\delta_\Lambda^{(m)} V$
is $m$ for $m \ge 1$,
and that in $\delta_\Omega^{(m)} \hskip1pt V$
is $m-1$ for $m \ge 1$.
Terms with the maximum number of $Q$ are written as
\begin{eqnarray}
\delta_\Lambda^{(m)} V
&=& g_{m+1} \, [\, V,\, (QV)^{m-1},\, Q \Lambda \,]
+ \mathcal{O}(Q^{m-1}) \,,
\label{delta-lambda-expansion} \\
\delta_\Omega^{(m)} \hskip1pt V
&=& h_{m+1}\, [\, \eta \hskip1pt \Omega,\, (QV)^{m-1},\, V \,]
+ \mathcal{O}(Q^{m-2}) \,,
\label{delta-omega-expansion}
\end{eqnarray}
for $m \ge 1$, where $g_{m+1}$ and $h_{m+1}$ are coefficients
to be determined.

\smallskip
Let us begin with the calculation of $(\delta_\Lambda S)_n$.
It follows from (\ref{S_n-expansion}),
(\ref{delta-lambda-expansion}), $S_2 = \mathcal{O}(Q)$,
and $\delta_\Lambda^{(0)} V = \mathcal{O}(Q)$ that
the maximum number of $Q$'s in $(\delta_\Lambda S)_n$
is $n+1$ for $n \ge 1$.
The terms in $(\delta_\Lambda S)_n$ which contain $(n+1)$ $Q$'s
and those which contain $n$ $Q$'s are given by
\begin{eqnarray}
(\delta_\Lambda S)_n
&=& \sum_{m=0}^{n} \delta_\Lambda^{(m)} S_{n-m+2}
\nonumber \\
&=& \Bigl[ \, g_{n+1} - (n+2) \, f_{n+2} \, \Bigr] \,
\{ \, Q \hskip1pt \eta V,\, V,\, (QV)^{n-1},\, Q \Lambda \, \}
\nonumber \\
&& {}+ 2 \sum_{m=0}^{n-2} \,
\biggl[ \, f_{n+2, \, m} - { \, n \, \choose \, m \, } \, f_{n+2} \,
\biggr] \,
\{ \, \eta V,\, (QV)^m,\, Q \Lambda,\,
[\, V,\, (QV)^{n-m-1} \,] \, \}
\nonumber \\
&& {}+ \sum_{m=1}^{n-1} \, \biggl[ \,
(m+2) \, g_{n-m+1} \, f_{m+2}
-2 \, { \, n \, \choose \, m \, } \, f_{n+2}
- f_{n+2, \, n-m-1} - f_{n+2, \, m-1} \, \biggr]
\nonumber \\
&& \qquad \qquad \times \,
\{ \, \eta V,\, (QV)^m,\,
[\, V,\, (QV)^{n-m-1},\, Q \Lambda \,] \, \}
\nonumber \\
&& {}+ \langle \, Q \hskip1pt \eta V,\,
\mathcal{O}(Q^{n-1}) \, \rangle
+ \mathcal{O}(Q^{n-1}) \,,
\label{delta-lambda-S}
\end{eqnarray}
where
${ \, n \, \choose \, m \,}$
is the binomial coefficient
${ \, n \, \choose \, m \, }
= \frac{n!}{m!
\, (n-m)!}
\,$.
Gauge invariance requires
\begin{eqnarray}
&& g_{n+1} - (n+2) \, f_{n+2} = 0 \,,
\nonumber \\
&& f_{n+2, \, m} - { \, n \, \choose \, m \, } \, f_{n+2} = 0 \,,
\label{lambda-equations}
\\
&& (m+2) \, g_{n-m+1} \, f_{m+2}
-2 \, { \, n \, \choose \, m \, } \, f_{n+2}
- f_{n+2, \, n-m-1} - f_{n+2, \, m-1} = 0 \,. \nonumber
\end{eqnarray}
These equations uniquely determine
$f_n$, $f_{n, \, m}$, and $g_{n}$ with $f_3 = 1/3! \,$:
\begin{equation}
f_n = \frac{1}{n!} \,, \quad
f_{n, \, m} = \frac{1}{n!} \, { \, n-2 \, \choose m } \,, \quad
g_{n} = \frac{1}{n!} \,.
\label{coefficients-from-Q}
\end{equation}

Let us next compute $(\delta_\Omega \hskip1pt S)_n$.
It follows from (\ref{S_n-expansion}),
(\ref{delta-omega-expansion}), $S_2 = \mathcal{O}(Q)$,
and $\delta_\Omega^{(0)} \hskip1pt V = \mathcal{O}(Q^0)$ that
the maximum number of $Q$'s in $(\delta_\Omega \hskip1pt S)_n$
is $n$ for $n \ge 1$.
The terms in $(\delta_\Omega \hskip1pt S)_n$
which contain $n$ $Q$'s
and the terms which contain $n-1$ $Q$'s are given by
\begin{eqnarray}
(\delta_\Omega \hskip1pt S)_n
&=& \sum_{m=0}^{n} \delta_\Omega^{(m)} \hskip1pt S_{n-m+2}
\nonumber \\
&=& \Bigl[ \, h_{n+1} - n \, (n+2) \, f_{n+2} \, \Bigr] \,
\{ \, Q \hskip1pt \eta V,\,
\eta \hskip1pt \Omega,\, (QV)^{n-1},\, V \, \}
\nonumber \\
&& \!\! + \sum_{m=0}^{n-2} \,
\biggl[ \, (n-m) \, f_{n+2, \, m}
- n \, { \, n-1 \, \choose m } \, f_{n+2} \, \biggr] \,
\{ \, \eta V,\,
[\, \eta \hskip1pt \Omega,\, (QV)^{n-m-1} \,] \, \,,\,
(QV)^m,\, V \, \}
\nonumber \\
&& \!\! + \sum_{m=0}^{n-3} \,
\biggl[ \, (m+1) \, f_{n+2, \, m+1}
- n \, { \, n-1 \, \choose m } \, f_{n+2} \, \biggr] \,
\{ \, \eta V,\,
[\, (QV)^{n-m-1} \,] \,,\, \eta \hskip1pt \Omega,\,
(QV)^m,\, V \, \}
\nonumber \\
&& \!\! + \sum_{m=1}^{n-1} \, \biggl[ \,
(m+2) \, h_{n-m+1} \, f_{m+2}
-2 \, n \, { \, n-1 \, \choose m } \, f_{n+2}
- (n-m) \, ( \, f_{n+2, \, n-m-1} + f_{n+2, \, m-1} \, ) \, \biggr]
\nonumber \\
&& \qquad \qquad \times \,
\{ \, \eta V,\,
[\, \eta \hskip1pt \Omega,\, (QV)^{n-m-1},\, V \,] \,,\,
(QV)^m \, \}
\nonumber \\
&& \!\! {}+ \langle \, Q \hskip1pt \eta V,\,
\mathcal{O}(Q^{n-2}) \, \rangle
+ \mathcal{O}(Q^{n-2}) \,.
\label{delta-omega-S}
\end{eqnarray}
Gauge invariance requires
\begin{equation}
\label{omega-equations}
\begin{split}
& h_{n+1} - n \, (n+2) \, f_{n+2} = 0 \,,
\\
& (n-m) \, f_{n+2, \, m}
- n \, { \, n-1 \, \choose m } \, f_{n+2}
= 0 \,,
\\
& (m+1) \, f_{n+2, \, m+1}
- n \, { \, n-1 \, \choose m } \, f_{n+2} = 0 \,,
\\
& (m+2) \, h_{n-m+1} \, f_{m+2}
-2 \, n \, { \, n-1 \, \choose m } \, f_{n+2}
- (n-m) \, ( \, f_{n+2, \, n-m-1} + f_{n+2, \, m-1} \, ) = 0 \,.
\end{split}
\end{equation}
These equations uniquely determine
$f_n$, $f_{n, \, m}$, and $h_{n}$ with $f_3 = 1/3! \,$:
\begin{equation}
f_n = \frac{1}{n!} \,, \quad
f_{n, \, m} = \frac{1}{n!} \, { \, n-2 \, \choose m } \,, \quad
h_n = \frac{n-1}{n!} \,.
\end{equation}
The coefficients $f_n$ and $f_{n, \, m}$
are in perfect agreement with those in (\ref{coefficients-from-Q}).
Since either one of the two sets of equations
(\ref{lambda-equations}) and (\ref{omega-equations}) is, by  itself,
an overdetermined system,  it is highly nontrivial
that there exists an action which respects gauge invariance
under both $\delta_\Lambda V$ and $\delta_\Omega \hskip1pt V$
up to this order.

\end{document}